\documentstyle[11pt,epsfig]{article}
\begin{document}

\def\no{{\noindent}}
\newcommand{\hsp}{\hspace*{3mm}}
\newcommand{\vsp}{\vspace*{3mm}}
\newcommand{\be}{\begin{equation}}
\newcommand{\ee}{\end{equation}}
\newcommand{\bea}{\begin{eqnarray}}
\newcommand{\eea}{\end{eqnarray}}
\newcommand{\bd}{\begin{displaymath}}
\newcommand{\ed}{\end{displaymath}}
\newcommand{\sgn}{~{\rm sgn}}
\newcommand{\IM}{{\rm Im}}
\newcommand{\RE}{{\rm Re}}
\newcommand{\bra}{\left\langle}
\newcommand{\ket}{\right\rangle}
\newcommand{\bigbra}{\left\langle}
\newcommand{\bigket}{\right\rangle}
\newcommand{\order}{{\cal O}}
\newcommand{\minus}{-}
\newcommand{\plus}{+}
\newcommand{\inn}{\!\cdot\!}
\newcommand{\hx}{\hat{x}}
\newcommand{\hy}{\hat{y}}
\newcommand{\hz}{\hat{z}}
\newcommand{\bv}{\mbox{\boldmath $v$}}
\newcommand{\bw}{\mbox{\boldmath $w$}}
\newcommand{\bJ}{\mbox{\boldmath $J$}}
\newcommand{\bB}{\mbox{\boldmath $B$}}
\newcommand{\bxi}{\mbox{\boldmath $\xi$}}
\newcommand{\bBS}{\mbox{{\boldmath $B$}}^*}

\renewcommand{\t}{T}
\newcommand{\th}{\hat{\t}}
\newcommand{\tav}{\overline{\t}}
\newcommand{\ta}{\tau}
\newcommand{\half}{\frac{1}{2}}
\newcommand{\bb}{\bBS}
\newcommand{\tl}{T^{\mu(\ell)}}
\newcommand{\tll}{T^{\mu(\ell')}}
\newcommand{\xl}{\bxi^{\mu(\ell)}}
\newcommand{\xll}{\bxi^{\mu(\ell')}}
\newcommand{\eg}{E_{\rm g}}
\newcommand{\egn}{E_{{\rm g}, 0}}
\newcommand{\ph}{\hat{P}_t}
\newcommand{\uu}{u}
\newcommand{\cc}{U}
\newcommand{\cct}{\widetilde{\cc}}
\newcommand{\rrt}{\widetilde{R}}

\title{\bf On-Line Learning with Restricted Training Sets:\\
An Exactly Solvable Case}

\author{\Large\bf H.C. Rae, P. Sollich and A.C.C. Coolen
\\[3mm]
Department of Mathematics \\
King's College, University of London\\
Strand, London WC2R 2LS, U.K.}

\maketitle

\begin{abstract}\noindent
We solve the dynamics of on-line Hebbian learning in large perceptrons
exactly, for the regime where the size of the training set scales
linearly with the number of inputs.  We consider both noiseless and
noisy teachers.  Our calculation cannot be extended to non-Hebbian
rules, but the solution provides a convenient and welcome benchmark
with which to test more general and advanced theories for solving the
dynamics of learning with restricted training sets.
\end{abstract}
\vsp

\vsp

\tableofcontents

\clearpage

\section{Introduction}

Considerable progress has been made in understanding the dynamics of
supervised learning in layered artificial neural networks through the
application of the methods of statistical mechanics.  A recent review
of work in this field is contained in~\cite{macecoolen}. For the most
part, such theories have concentrated on systems where the training
set is much larger than the number of weight updates.  In such
circumstances the probability that any given question will be repeated
during the training process is negligible and it is possible to assume
for large networks, via the central limit theorem, that their local
field distribution is always Gaussian.  In this paper we consider {\it
restricted training sets}; we suppose that the size $p$ of the
training set scales linearly with $N$, the number of inputs. As a
consequence the probability that a question will reappear during the
training process is no longer negligible, the assumption that the
local fields have Gaussian distributions is not tenable, and it is
clear that correlations will develop between the weights and the
questions in the training set as training progresses.  In fact, the
non-Gaussian character of the local fields should be a {\it
prediction} of any satisfactory theory of learning with restricted
training sets, as this is clearly demanded by numerical simulations.

Several authors~\cite{horner,KroghHertz,%
sollichbarber1,sollichbarber2,coolensaad,coolensaad2} have discussed
learning with restricted training sets but constructing a general
theory is difficult. A simple model of learning with restricted
training sets which can be solved {\it exactly } is therefore
particularly attractive and provides a yardstick against which more
difficult and sophisticated general theories can, in due course, be
tested and compared.  We show how this can be accomplished for on-line
Hebbian learning in perceptrons with restricted training sets and we
obtain exact solutions for the generalization error, the training
error and the field distribution for a class of noisy teacher networks
and student networks with arbitrary weight decay. We work out in
detail the two particular but representative cases of output noise and
Gaussian weight noise.  Our theory is found to be in excellent
agreement with numerical simulations and our predictions for the
probability density of the student field are a striking confirmation
of them, making it clear that we are indeed dealing with local fields
which are non-Gaussian. An outline of our results is to appear in the
conference proceedings~\cite{NIPS_Hebbonline_ref}.

\section{Definitions and Explicit Microscopic Expressions}

We study on-line learning in a student perceptron $S$, which tries to
learn a task defined by a noisy teacher perceptron $T$. The student
input-output mapping is specified by a weight vector $\bJ$ according
to
\bd
S:\{-1,1\}^N\rightarrow \{-1,1\} 
\qquad
S(\bxi)=\sgn[{\bJ}\cdot\bxi]\ .
\ed
For a given $\bJ$, this is a deterministic mapping from binary inputs
to binary outputs.  The teacher output $T(\bxi)$, on the other hand,
is stochastic. In its most general form, it is determined by the
probabilities $P(\t=\pm 1|\bxi)$. These are related to the {\em
average} teacher output $\tav(\bxi)$ for a given input $\bxi$ by
\be
P(\t=\pm 1|\bxi) = \half[1\pm \tav(\bxi)], \qquad \mbox{or} \qquad
P(\t|\bxi) = \half[1+\t\tav(\bxi)]\ .
\label{P_tav_relation}
\ee
To ensure that this noisy teacher mapping can be thought of as the
corrupted output of an underlying `clean' perceptron with weights
$\bb$, we make the mild assumption that the average teacher output
can be written in the form
\be
\tav(\bxi)=\ta(y), \qquad y=\bb\cdot\bxi
\label{tau_definition}
\ee
with some function $\ta(y)$. In other words, the noise process
preserves, on average, the perceptron structure of the teacher. The
uncorrupted teacher weight vector is taken to be normalized such that
$(\bb)^2=1$, with each component $B^*_i$ of $\order (N^{-\frac
{1}{2}})$. We also assume that inputs are sampled randomly from a
uniform distribution\footnote{%
This choice of input distribution is not critical. In fact, any other
distribution with $\bra\xi_i\ket=0$ and
$\bra\xi_i\xi_j\ket=\delta_{ij}$ will give results identical to the
ones for the present case in the limit $N\to\infty$. Examples would be
real-valued inputs with either a Gaussian distribution with zero mean
and unit variance, or a uniform distribution over the hypersphere
$\bxi^2=N$. Likewise, we only actually require that
assumption~(\protect\ref{tau_definition}) should hold with probability
one (i.e., for almost all inputs) in the limit $N\to\infty$. }
on $\{-1,1\}^N$. Typical values of the (uncorrupted) 'teacher field'
$y$ are then of $\order(1)$; in the thermodynamic limit $N\rightarrow
\infty$ that we will be interested in, $y$ is Gaussian with zero mean
and unit variance.

The class of noise processes allowed by~(\ref{tau_definition}) is
quite large and includes the standard cases of output noise and
Gaussian weight noise that are often discussed in the literature. For
output noise, the sign of the clean teacher output $\sgn(y)$ is
inverted with probability $\lambda$, i.e.,
\be
P(\t|\bxi) = (1-\lambda)\ \theta(\t y) + \lambda\theta(-\t y),
\qquad
\ta(y)=(1-2\lambda)\sgn(y)
\label{flip}
\ee
For Gaussian weight noise, the teacher output is produced from a corrupted
teacher weight vector $\bB$. The corrupted weights $\bB$ differ from $\bb$
by the addition of Gaussian noise of standard deviation
$\Sigma/\sqrt{N}$ to each component, i.e.,
\be
P(\bB)=\left[\frac {N}{2\pi\Sigma^2}\right]^{N/2}
\exp\left(-\frac{N}{2\Sigma^2}(\bB -\bBS)^2\right).
\label{gauss}
\ee
The scaling with $N$ here is chosen to get a sensible result in the
thermodynamic limit (corrupted and clean weights clearly need to be of
the same order). The corrupted teacher field is then
$z=\bB\cdot\bxi=y+\Delta$, with $\Delta$ a Gaussian random variable
with zero mean and variance $\Sigma^2$, and hence
\be
\ta(y) = \bra \sgn(y+\Delta) \ket_{\Delta} = {\rm erf}(y/\sqrt{2}\Sigma).
\label{gauss2}
\ee
In the numerical examples presented later, we will focus on the above
two noise models.  But our analytical treatment applies to any teacher
that is compatible with the assumption~(\ref{tau_definition}). This
covers, for example, the more complex cases of `reversed wedge'
teachers (where $\ta(y)=\sgn(y)$ for $|y|>d$ and $\ta(y)=-\sgn(y)$
otherwise, $d$ being the wedge `thickness') and noisy generalizations
of these.

Our learning rule will be the on-line Hebbian rule, i.e.
\be
\bJ (\ell \plus 1)= \left(1-\frac {\gamma }{N}\right) \bJ(\ell)
+\frac {\eta}{N}\ \xl\tl
\label{b1}
\ee
where the non-negative parameters $\gamma$ and $\eta$ are the weight
decay and the learning rate, respectively. Learning starts from an
initial set of student weights $\bJ_0\equiv\bJ(0)$, for which we
assume (as for the teacher weights) that $J_i(0)\!=\!\order (N^{-\frac
{1}{2}})$. At each iteration step $\ell$ a training example,
comprising an input vector $\xl$ and the corresponding teacher output
$\tl$, is picked at random (with replacement) from the {\em training
set} $D$. This training set consists of $p=\alpha N$ examples,
$D=\{(\bxi^\mu, \t^\mu),$ $\mu =1\ldots p\}$; it remains unchanged
throughout the learning process. Each training input vector $\bxi^\mu$
is assumed to be randomly drawn from $\{-1,1\}^N$ (independently of
other training inputs, and of $\bJ_0$ and $\bb$), and the output
$\t^\mu=\t(\bxi^\mu)$ provided by the noisy teacher. We call this kind of
scenario `consistent noise': To each training input corresponds
a single output value which is produced by the teacher once and for
all before learning begins; the teacher is {\em not} asked to produce
new noisy outputs each time a training input is selected for a
weight update.

There are two sources of randomness in the above scenario. First of
all there is the random realization of the `path' $\Omega =\{\mu(0),
\mu(1), \mu(\ell), \ldots \}$.  This is simply the dynamic randomness
of the stochastic process that gives the evolution of the vector
$\bJ$; it arises from the random selection of examples from the
training set.  Averages over this process will be denoted as $\bra
\ldots \ket$. Secondly there is the randomness in the composition of
the training set.  We will write averages over all training sets as
$\bra \ldots \ket _{\rm {sets}}$. We note that
\bd
{\bra f(\xl,\tl)\ket }=\frac{1}{p}\sum_{\mu =1}^p f({\bxi}^\mu,\t^\mu)
\qquad
\mbox{(for all $\ell$)}
\ed
and that averages over all possible realizations of the training set
are given by
\begin{eqnarray*}
\lefteqn{ 
\bra f[(\bxi ^1,\bB^1),(\bxi ^2,\bB^2),\ldots,
(\bxi^p,\bB^p)]\ket_{\rm {sets}} = } \nonumber\\
& = & 
\sum_{\bxi ^1}\,\sum_{\bxi ^2}\ldots\,\sum_{\bxi^p}
\left(\frac{1}{2^{N}}\right)^p
\sum_{\t^1, \ldots, \t^p=\pm 1} \left[\prod _{\mu=1}^p P(\t^\mu|\bxi^\mu)
\right] \ f[(\bxi ^1,\bB^1),(\bxi ^2,\bB^2),\ldots,(\bxi
^p,\bB^p)]
\end{eqnarray*}
where $\bxi^{\mu} \in \{-1,1\}^N$.

Our aim is to evaluate the performance of the on-line Hebbian learning
rule~(\ref{b1}) as a function of the number of training steps
$m$. This calculation becomes tractable in the thermodynamic limit
$N\to\infty$; the appropriate time variable in this limit is $t=m/N$.
Basic quantities of interest are the generalization error and the
training error. The generalization error, which we choose to measure
with respect to the {\em clean} teacher, is the probability of student
and (clean) teacher producing different outputs on a randomly chosen
test input. Hence $\eg=\bra
\theta[-(\bJ\inn\bxi)(\bB^*\inn\bxi)]\ket_{\bxi}$, with the usual
result
\be
\eg=\frac {1}{\pi} \arccos\left(\frac{R}{\sqrt{Q}}\right).
\label{aa4}
\ee
Here $Q=\bJ^2$ is the squared length of the student
weight vector, and $R=\bBS \inn \bJ$ its overlap with the teacher
weights. These are our basic scalar observables. The
training error $E_{\rm t}$ is the fraction of errors that the students
makes on the training set, i.e., the fraction of training outputs that
are predicted incorrectly. It is given by
\bd
E_{\rm t} = \int\! dx \, \sum_{\t=\pm 1} P(x,\t)\, \theta(-\t x)
\ed
where $P(x,\t)$ is the joint distribution of the student fields $x=\bJ
\inn \bxi$ and the teacher outputs $\t$ over the training set. Because
the teacher outputs depend on the teacher fields, according to
$P(\t|y)=\half[1 + \t \ta(y)]$, it is useful to include the latter
and to calculate the distribution $P(x,y,\t)$; we will see later that
this also leads to a rather transparent form of the result. Formally,
the joint field/output distribution is defined in the obvious way,
\be
P(x,y,\t)=\frac{1}{p}\sum_{\mu=1}^p\ 
\delta(x- \bJ\inn\bxi^\mu)\ \delta(y-\bBS\inn\bxi^\mu)\
\delta_{\t,\t^\mu}\ .
\label{aa2}
\ee

For infinitely large systems, $N\to\infty$, one can prove that the
fluctuations in mean-field observables such as $\{Q,R,P(x,y,\t)\}$,
due to the randomness in the dynamics, will
vanish~\cite{coolensaad}. Furthermore one assumes, with convincing
support from numerical simulations, that for $N\to\infty$ the
evolution of such observables, when observed for different random
realizations of the training set, will be reproducible (i.e., the
sample-to-sample fluctuations will also vanish, which is called
`self-averaging'). Both properties are central ingredients of all
current theories.  We are thus led to the introduction of averages of
our observables, both with respect to the dynamical randomness and
with respect to the randomness in the training set (always to be
carried out in precisely this order):
 \be
Q(t)=\lim_{N\to\infty}\bra\, \bra Q\ket\,\ket_{\rm sets}
\qquad
R(t)=\lim_{N\to\infty}\bra \,\bra R\ket\,\ket_{\rm sets}
\label{eq:averages1}
\ee
\be
P_t(x,y,\t)=\lim_{N\to\infty}\bra\, \bra P(x,y,\t)\ket \,\ket_{\rm sets}
\label{eq:averages2}
\ee
The large $N$-limits here are taken at constant $t$ and $\alpha$,
i.e., with the number of weight updates and the number of training
examples scaling as $m=Nt$ and $p=N\alpha$, respectively.

Iterating the learning rule~(\ref{b1}), we find an explicit expression
for the student weight vector after $m$ training steps:
\be
\bJ (m)= \sigma^{m} \bJ_0 \ +\ \frac {\eta
}{N}\sum_{\ell =0}^{m-1}\sigma^{m-\ell -1}\xl\,\tl
\label{b2}
\ee
where
\bd
\sigma =1\minus\frac{\gamma}{N}\ .
\ed
Eq.~(\ref{b2}) will be the natural starting point for our
calculation. We will also frequently encounter averages of the form
\[
\bra\bv\cdot\bxi\,\t(\bxi)\ket_{\bxi,\t}
\]
which we now calculate.  The average over $\t$ is trivial and, using
assumption~(\ref{tau_definition}), gives
$\langle\bv\cdot\bxi\,\ta(\bBS\cdot\bxi)\rangle_{\bxi}$. Provided
all components of the vector $\bv$ are of the same order,
$v=\bv\cdot\bxi$ and $y=\bBS\cdot\bxi$ become zero mean Gaussian
variables for $N\to\infty$ with $\bra vy\ket=\bv\cdot\bBS$ and $\bra
y^2\ket=(\bBS)^2=1$. By averaging over $v$ first for fixed $y$, we
thus obtain the desired result
\be
\bra\bv\cdot\bxi\,\t(\bxi)\ket_{\bxi,\t} = \rho\, \bv\cdot\bBS,
\qquad \rho=\bra y\ta(y)\ket = \int\!\! Dy\, y \, \ta(y)
\label{rho}
\ee
with the familiar short-hand $Dy=(2\pi)^{-\frac{1}{2}}e^{-z^2/2}dy$.
Using~(\ref{flip}) and~(\ref{gauss2}), one thus finds for the
proportionality constant $\rho$
\bea
\rho & = & \sqrt {\frac {2}{\pi }}\ (1\minus 2\lambda )
\qquad 
\mbox{(output noise)}
\label{888}
\\
\rho & = & \sqrt {\frac {2}{\pi }}\frac {1}{\sqrt {1+{\Sigma}^2}}
\qquad
\mbox{(Gaussian weight noise)}
\label{999}
\eea
for the two noise models that we will consider in some detail.

\section{Simple Scalar Observables}

It is a simple matter to calculate the values of $Q$ and $R$ after
$m$ learning steps, using~(\ref{b1}). For $Q$, we find
%
\bd
Q = \sigma^{2m}\bJ_0^2
+\frac {2\eta }{N}\sum _{\ell =0}^{m-1}
\sigma^{2m-\ell -1}{\bJ _0}\inn\xl\,\tl
+ \frac {\eta ^2}{N^2}\sum_{\ell ,\ell'=0}^{m-1}
\sigma^{m-\ell -1} \sigma^{m-\ell'-1} \xl\cdot\xll \, \tl \tll
\ed
%
We now average both with respect to dynamical (or path) randomness and
with respect to the randomness in the training set, and take the limit
$N\rightarrow\infty$ at constant learning time $t=m/N$
(see~(\ref{eq:averages1})). Separating out the terms with $\ell=\ell'$
from the double sum, and using~(\ref{rho}), we obtain
\bea 
Q(t) & = & e^{-2\gamma t}Q_0
+2\eta\rho R_0\lim _{N\rightarrow \infty }
\frac {1}{N}\sum _{\ell =0}^{tN}\sigma^{2tN-\ell} 
+\eta ^2\lim _{N\rightarrow \infty}\frac{1}{N}\sum _{\ell =0}^{tN}
\sigma^{2tN-2\ell }
\nonumber\\
& & +\ 
\eta ^2 \lim _{N\rightarrow \infty } \frac {1}{N^2}\sum _{\ell
\neq \ell'}
\sigma^{tN-\ell }
\sigma^{tN-\ell'}
\bra\bra\xl\cdot\xll \,\tl \tll\ket\ket_{\rm sets}.
\nonumber
\eea
Here $Q_0=\bJ_0^2$ and $R_0=\bJ_0\cdot\bBS$ are the squared length and
overlap of the initial student weights, respectively. After averaging
over the dynamical randomness, the average in the last term becomes
$(1/p^2) \sum_{\mu,\nu=1}^p \bra\bxi^\mu \cdot \bxi^\nu \, \t^\mu
\t^\nu\ket_{\rm sets}$. The terms with $\mu=\nu$ each contribute
$(\bxi^\mu)^2=N$ to this sum; the others make a contribution of
$\rho^2$ each, as one finds by applying~(\ref{rho}) twice.  Assembling
everything, we have
\be
Q(t) = e^{-2\gamma t}Q_0
+2\rho R_0\frac {\eta}{\gamma } e^{-\gamma t}(1\minus e^{-\gamma t})
+\frac {\eta ^2}{2\gamma }(1\minus e^{-2\gamma t})
+\frac{\eta ^2}{\gamma ^2}
\left(\frac {1}{\alpha }\plus{\rho}^2\right)
(1\minus e^{-\gamma t})^2
\label{b4}
\ee
where $\rho$ is given by equations~(\ref{888}, \ref{999}) in the
examples of output noise and Gaussian weight noise, respectively, and
more generally by~(\ref{rho}). In a similar manner we find that
\bea
R(t) & = & \lim _{N\rightarrow \infty}\ 
\sigma^{tN}R_0
+\frac {\eta }{N}\sum_{\ell=0}^{tN}
\sigma^{tN-\ell } \bra \bra\ 
\bBS\cdot\xl\,\tl \ \ket \ket_{\rm {sets}}
\nonumber\\
& = & e^{-\gamma t}R_0+
\frac {\eta \rho}{\gamma} (1-e^{-\gamma t})
\label{b3}
\eea
We note in passing that our calculations easily generalize to the case
of a variable learning rate $\eta (t)$.  Sums such as $\frac {\eta
}{N}\sum _{\ell =0}^{tN}\sigma^{tN-\ell }$ would simply be replaced by
$\frac {1}{N}\sum _{\ell =0}^{tN} \sigma^{tN-\ell }\eta (\ell
/N)$. Using $\sigma^{tN-\ell}=(1-\gamma /N )^{tN-\ell }=\exp[-\gamma t
+\gamma \ell/N+\order (1/N)]$ we see that
\bd
\lim_{N\rightarrow \infty }\frac {1}{N}\sum _{\ell =0}^{tN}
\sigma^{tN-\ell }\eta (\ell /N)
=\int _0^t\! ds~e^{-\gamma(t-s)}\eta (s)
\ed
which reduces to the familiar result in the case when $\eta $ is
constant.  Other sums involving a variable learning rate can be
treated in similar fashion.

\begin{figure}[t]
\begin{center}
\vspace*{-0.8cm}
\epsfig{file=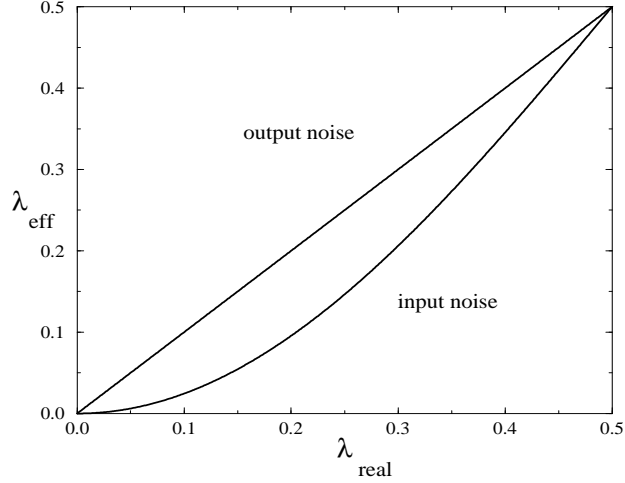}
\end{center}
\vspace*{-0.6cm}
\caption{Relation between the real output error probability
$\lambda_{\rm real}$ and the effective output error probability
$\lambda_{\rm eff}$ for noisy teachers. For output noise the two are
identical. One observes that, for the same `real' noise level, weight
noise is significantly less disruptive to the learning process than
output noise.}
\label{fig:lambdas}
\end{figure}

The generalization error follows directly from the above results
and~(\ref{aa4}); its asymptotic value is
\be
\lim _{t\rightarrow \infty } \eg(t)=\frac {1}{\pi }
\arccos \left(\frac {\rho}{\sqrt {\gamma/2
+1/\alpha + \rho^2}}\right)
\label{eginf}
\ee
More generally, one sees from~(\ref{b4},\ref{b3}) that (for
$N\to\infty$) all noisy teachers with the same $\rho$ will give the
same generalization error at any time $t$. This is true, in
particular, of output noise and Gaussian weight noise when their
respective parameters $\lambda$ and $\Sigma$ are related by
$1-2\lambda =(1\plus {\Sigma }^2)^{-\frac{1}{2}}$.  More generally,
one can use~(\ref{888}) to associate, with any type of teacher noise
obeying our basic assumption~(\ref{tau_definition}), an effective
output noise parameter $\lambda_{\rm{eff}}$ given by
\be
1-2\lambda_{\rm{eff}} = \sqrt{\frac{\pi}{2}}\,\rho = 
\sqrt{\frac{\pi}{2}}\,\bra y\ta(y)\ket
\label{lambda_eff}
\ee
Note, however, that this effective teacher error probability $\lambda
_{\rm{eff}}$ will in general not be identical to the {\em real}
teacher error probability $\lambda _{\rm{real}}$. The latter is
defined as the probability of an incorrect teacher output for a random
input, $\lambda _{\rm{real}}=$ $\bra P(\t=
-\sgn(\bBS\cdot\bxi)|\bxi) \ket_{\bxi}$. Using~(\ref{P_tav_relation}),
this can be rewritten as $\lambda _{\rm{real}} =
\bra\half[1-\sgn(\bBS\cdot\bxi)\tav(\bxi)]\ket_{\bxi}$, and
with~(\ref{tau_definition}) one obtains
\be
1-2\lambda _{\rm{real}} = \bra\sgn(y)\ta(y)\ket.
\label{lambda_real}
\ee
Comparing with~(\ref{lambda_eff}), one sees that in the effective
error probability that is relevant to our Hebbian learning process,
errors for inputs with large teacher fields $y$ are weighted more
heavily than in the real error probability. For output noise, this is
irrelevant because the probability of an incorrect teacher error is
independent of $y$, and $\lambda _{\rm{real}}$ and $\lambda
_{\rm{eff}}$ are therefore identical. For Gaussian weight noise, on
the other hand, errors are most likely to occur near the decision
boundary of the teacher ($y=0$). These are suppressed by the weighting
in the effective error probability, and so $\lambda
_{\rm{eff}}<\lambda_{\rm real}$. Explicitly, one finds in this case
$\lambda_{\rm real} = \frac{1}{\pi}\arctan\Sigma$, and
from~(\ref{999}), $\lambda_{\rm eff}=\frac {1}{2}[1 - (1+\Sigma
^2)^{-1/2}]$; the relation between effective and real error
probabilities for Gaussian weight noise (see figure \ref{fig:lambdas})
is therefore
\bd
\lambda _{\rm {eff}}=\frac {1}{2}\bigl [1-\cos (\pi \lambda _{\rm
{real}}\bigr )]={\sin }^2(\pi \lambda _{\rm {real}}/2).
\ed

We now briefly consider whether the generalization error $\eg(t)$ can
have a minimum at a finite time $t$, i.e., whether overtraining can
occur in our problem.  After a straightforward but tedious calculation
we find that $\eg(t)$ as given by~(\ref{aa4},\ref{b4},\ref{b3}) is
stationary at the point $t=t^*$, where
\be
t^*=\frac {1}{\gamma }\log \left[\frac
{-2\gamma\eta^{-1}\rho Q_0\sin^2(\pi \egn)
-2Q_0^{1/2}\alpha^{-1}\cos(\pi\egn)+\eta\rho}
{\eta\rho-Q_0^{1/2}(2/\alpha+\gamma)\cos(\pi\egn)}\right]
\label{b5}
\ee
Here $\egn\equiv \eg(0)$ is the initial generalization performance of
the student.  It turns out that $\eg(t)$ has a {\it minimum } at $t^*$
if the numerator of the logarithm in equation (\ref{b5}) is negative.
Of course, $t^*$ must be real and {\it positive} -- which demands that
the denominator of the logarithmic term in (\ref{b5}) be negative, and
that the numerator be less than the denominator.  This implies that
$\eg(t)$ will have a minimum at $t^*$ if
\be
\alpha < \frac {2Q_0^{1/2}\cos(\pi \egn)}{\eta\rho-\gamma
Q_0^{1/2}\cos(\pi\egn)}
\qquad
\mbox{and}
\qquad
\eta \cos(\pi\egn) < 2Q_0^{1/2}\rho\, \sin^2(\pi\egn)
\label{b6}
\ee
In writing~(\ref{b6}) we have made the reasonable assumption that
$\egn\in [0,\frac{1}{2}]$, corresponding to an initial performance no
worse than random guessing. When the conditions~(\ref{b6}) are
satisfied the generalization error has a minimum at $t^*$ and
overtraining occurs for $t>t^*$.  However, in practice this phenomenon
does not appear to be of great significance, with typically $t^*<1$.

\section{Joint Field Distribution}
\label{sec:distribution}

The calculation of the average of the joint field distribution
starting from equation~(\ref{eq:averages2}) is more difficult than
that of the scalar observables.  It is convenient to work in terms of
the characteristic function
\be
\ph(\hx,\hy,\th)=\bra e^{-i(\hx x+\hy y+\th\t)}\ket_{P_t(x,y,\t)} \ ;
\label{char_function}
\ee
using equations~(\ref{aa2},\ref{eq:averages2},\ref{b2}), we then find that
\begin{eqnarray}
\ph(\hx,\hy,\th) &=& \lim _{N\rightarrow \infty} 
\bra \frac {1}{p}\sum_{\mu=1}^p
\exp[-i(\hx{\sigma}^{tN}\bJ_0\cdot\bxi^\mu+\hy\bBS\cdot\bxi^\mu+\th \t^\mu)]
\right.
\nonumber\\
& & \times \left.
\bra \exp\left(-\frac{i\eta\hx}{N}\sum_{\ell=0}^{tN}
{\sigma}^{tN-\ell}\, \bxi^\mu\cdot\xl\,\tl \right)\ket\,\ket_{{\rm sets}}
\label{intermediate}
\end{eqnarray}
Performing the path average gives
\bd
\bra \exp\left(-\frac{i\eta\hx}{N}\sum_{\ell=0}^{tN}{\sigma}^{tN-\ell}
\bxi^\mu\cdot \xl\,\tl \right)\ket
=\prod_{\ell =0}^{tN}
\left[ 
\frac{1}{p}\sum_{\nu =1}^p \exp\left(-\frac {i\eta \hx
}{N}{\sigma}^{tN-\ell}\bxi^\mu \cdot \bxi^\nu \, \t^\nu \right)
\right]
\ed
After substitution of this result into~(\ref{intermediate}), only a
training set average remains.  Once this has been carried out, all
terms in the sum over $\mu$ will be exactly equal. Anticipating this
by setting $\mu=1$, we get
\begin{eqnarray}
\ph(\hx,\hy,\th) &=& \lim _{N\rightarrow \infty} 
\bra \rule[-3mm]{0mm}{9mm}
\exp[-i(\hx{\sigma}^{tN}\bJ_0\cdot\bxi^1+\hy\bBS\cdot\bxi^1+\th \t^1)]
\right.
\nonumber\\
& &\times \left.
\prod_{\ell =0}^{tN}
\left[
\frac{1}{p}\sum_{\nu =1}^p \exp\left(-\frac {i\eta\hx}{N}\sigma^{tN-\ell}
\bxi^1 \cdot \bxi^\nu \, \t^\nu \right) \right]
\ket_{\rm sets}\ .
\label{c3}
\end{eqnarray}
Consider now the product $S=\prod _{\ell =0}^{tN}[\ldots]$. The
$\nu\!=\!1$ term of the sum in square brackets needs to be
treated separately because $\bxi^1\cdot\bxi^1=N$. For $\nu>1$, on the
other hand, the products $\bxi^1\cdot\bxi^\nu$ are overlaps between
{\em different} input vectors and therefore only of
$\order(\sqrt{N})$; the rescaled overlaps $v_\nu=\bxi^1\inn \bxi
^\nu/\sqrt {N}$ are of $\order(1)$. In the sum over $\nu>1$ in
\bd
\log S = \sum _{\ell =0}^{tN} \log\left[
\frac {1}{p} \exp\left(-i\eta \hx {\sigma}^{tN-\ell}\t^1\right)
+ \frac {1}{p} \sum _{\nu >1}
\exp\left(-\frac{i\eta \hx}{\sqrt{N}} {\sigma}^{tN-\ell}
v_\nu\t^\nu\right)
\right]
\ed
the exponential therefore has an argument of $\order(N^{-1/2})$ and
can be Taylor expanded. Terms up to $\order(1/N)$ (i.e., up to second
order) need to be retained because of the sum over the $\order(N)$
values of $\ell$, and so
\begin{eqnarray*}
\lefteqn{\log S =}
\\
& = & \sum _{\ell =0}^{tN} \log\left[
\frac {1}{p} 
\exp\left(-i\eta \hx {\sigma}^{tN-\ell}\t^1\right)
+ \frac {p-1}{p} + \frac {1}{p} \sum _{\nu >1}
\left(-\frac{i\eta \hx}{\sqrt{N}} {\sigma}^{tN-\ell} v_\nu\t^\nu
-\frac{\eta^2 \hx^2}{2N} {\sigma}^{2tN-2\ell} v_\nu^2
\right)
\right]\\
& = & \frac {1}{p} \sum_{\ell =0}^{tN} \left[
\exp\left(-i\eta \hx {\sigma}^{tN-\ell}\t^1\right)-1
 -i\eta \hx {\sigma}^{tN-\ell} \frac{1}{\sqrt{N}} \sum _{\nu >1} v_\nu\t^\nu
-\half\eta^2 \hx^2 {\sigma}^{2tN-2\ell} \frac{1}{N} \sum _{\nu >1} v_\nu^2
\right]
\end{eqnarray*}
where contributions of $\order(N^{-1/2})$ have been discarded.
Transforming the first sum over $l$ into an integral over time (by considering
appropriate Riemann sums), we then obtain
\be
\log S =\chi (\hx \t^1) -\frac{i\eta \hx u_1}{\gamma}(1\minus e^{-\gamma t})
-\frac {\eta ^2{\hx }^2 u_2}{4\gamma}(1\minus e^{-2\gamma t})
\label{aa5}
\ee
where
\be
\chi (w)=\frac {1}{\alpha }\int _0^t\! ds\,
\left\{\exp\left[-i\eta w e^{-\gamma(t-s)}\right]\minus 1\right\}
\label{chi_def}
\ee
and
\bd
u_1=\frac {1}{\alpha \sqrt {N}}\sum _{\nu >1}v_\nu \t^\nu
\qquad
u_2=\frac{1}{p}\sum _{\nu>1}v_\nu^2\ .
\ed
Further progress requires considering the statistics of the random
variables $u_1$ and $u_2$. For $N\to\infty$, the $v_\nu$ are
independent Gaussian variables with zero mean and unit variance. By
the central limit theorem, $u_2$ therefore has fluctuations of
$\order(N^{-1/2})$ and can be replaced by its average $\bra u_2
\ket=1$ in the thermodynamic limit. Similarly, because the products
$v_\nu\t^\nu$ are uncorrelated for different $\nu$, $u_1$ becomes
Gaussian in this limit. Using~(\ref{rho}), its mean and variance can
be calculated as
\begin{eqnarray*}
\bra u_1\ket & = &
\frac {1}{\alpha \sqrt{N}}\sum _{\nu >1} \bra v_\nu \t^\nu \ket = 
\frac{p-1}{\alpha N} \bra \bxi^1\cdot\bxi\, \t(\bxi)\ket =
\rho\,\bBS\cdot\bxi^1 + \order(N^{-1})
\\
\bra(\Delta u_1)^2\ket & = &
\frac{p-1}{\alpha N} \left[ \bra v_\nu^2 \ket - \bra v_\nu
\ket^2\right] =
\frac{1}{\alpha}\left[1-\order(N^{-1})\right]
\end{eqnarray*}
We conclude that, for large $N$, $u_1=\rho\,\bBS\inn \bxi^1 \plus
\alpha ^{-1/2}\hat {u}$, where $\hat {u}$ is a unit variance Gaussian
random variable with mean zero.  We are now in a position to average
$S$ as given by~(\ref{aa5}) over all realizations of
$\{(\bxi^\nu,\t^\nu), \nu>1\}$, with the result
\bd
\bra S\ket = \exp\left[\chi (\hx \t^1) - 
\frac{i\eta \hx \rho \bBS\inn \bxi^1}{\gamma}(1\minus e^{-\gamma t})
-\frac{\eta^2 \hx^2}{2\alpha\gamma^2}(1\minus e^{-\gamma t})^2
-\frac{\eta^2 \hx^2}{4\gamma}(1\minus e^{-2\gamma t})\right]
\ed
Inserting this into equation~(\ref{c3}) for the characteristic
function, we are left with a final average over $\bxi^1$ and $\t^1$,
with the former entering only through the fields
$\uu=\bJ_0\cdot\bxi^1$ and $y^1=\bBS\cdot\bxi^1$:
\begin{eqnarray}
\ph(\hx,\hy,\th) & = & \bra 
\exp\left[-i(\hx e^{-\gamma t} \uu+\hy y^1+\th \t^1)
+ \chi (\hx \t^1) - 
\frac{i\eta \hx \rho y^1}{\gamma}(1\minus e^{-\gamma t})
\right. \right.
\nonumber\\
& &
\left. \left.
-\frac{\eta^2 \hx^2}{2\alpha\gamma^2}(1\minus e^{-\gamma t})^2
-\frac{\eta^2 \hx^2}{4\gamma}(1\minus e^{-2\gamma t})\right]
\ket_{\uu,y^1,\t^1}
\label{almost_there}
\end{eqnarray}
We now observe that $\t^1$ only depends on $y^1$, but not on $\uu$;
correspondingly, $\uu$ is independent of $\t^1$ if $y^1$ is given. For
large $N$, the two fields $\uu$ and $y^1$ are zero mean Gaussian
random variables with $\bra \uu^2 \ket = Q_0$, $\bra \uu y^1 \ket =
R_0$ and $\bra (y^1)^2 \ket = 1$. The average of the $\uu$-dependent
factor in~(\ref{almost_there}), for given $y^1$, is therefore
\bd
\bra \exp\left(-i\hx e^{-\gamma t}\uu\right) \ket_{\uu|y^1} = 
\exp\left[-i\hx e^{-\gamma t} R_0y^1 - \half\hx^2 e^{-2\gamma t}
(Q_0-R_0^2)
\right]
\ed
Inserting this into~(\ref{almost_there}), and
using~(\ref{b4},\ref{b3}), one finds that the terms in the exponential
which are linear in $\hx$ combine to a term proportional to $R(t)$,
whereas the quadratic terms in $\hx$ conspire to give a contribution
proportional to $Q(t)-R^2(t)$:
\be
\ph(\hx,\hy,\th) = \bra 
\exp\left[
 - i \hy y^1 - i \th \t^1 + \chi (\hx \t^1) - \half\hx^2(Q-R^2) - i R\hx y^1
\right]\ket_{y^1,\t^1}
\label{almost2}
\ee
Finally, we recast this result in terms of the conditional
distribution of $x$, given $y$ and $\t$. To do this, first note that
the distribution of $y^1$ and $\t^1$ that is to be averaged over on
the right hand side of~(\ref{almost2}) is just the distribution of the
teacher field $y$ and the teacher output $\t$ over the training
set. We rename them appropriately and write out the
definition~(\ref{char_function}) of the characteristic function
on the left hand side:
\begin{eqnarray*}
\lefteqn{
\int\! dx\, dy \sum_{\t=\pm 1} \exp\left[- i \hy y - i \th \t - i \hx x\right]
\, P_t(x|y,\t) \, P(y,\t)=
}\\
& & \int\! dy \sum_{\t=\pm 1} \exp\left[
 - i \hy y - i \th \t + \chi (\hx \t) - \half\hx^2(Q-R^2) - i R\hx y
\right] \, P(y,\t).
\end{eqnarray*}
Equality for all $\hy$ and $\th$ implies that 
\bd
\int \! dx\, \exp(-i \hx x)\, P_t(x|y,\t) = 
%
%
\exp\left[ \chi (\hx \t) - \half\hx^2(Q-R^2) - i R\hx y\right]
\ed
and hence our final result\footnote{%
Equation~(\ref{PS1}) can also be derived by using Fourier transforms
to obtain $P_t(x,y,\t)$ from~(\ref{almost2}), and then dividing by
$P(y,\t)$.
}
\be
P_t(x|y,\t) = \int \frac{d\hx}{2\pi}\exp\left[i \hx (x-Ry) +
\chi (\hx \t) - \half\hx^2(Q-R^2)\right]
\label{PS1}
\ee
which is remarkably simple. In particular, we note that in this
conditional distribution of $x$, the noise properties enter only
through the parameter $\rho$; in fact, they only affect the factor
$\exp(-i\hx R y)$, while both $Q-R^2$ and $\chi(\hx \t)$ are actually
independent of $\rho$. Eq.~(\ref{PS1}) also shows that the dependence
of the student field on $y$ and $\t$ can be written in the simple form
\bd
x=Ry + \Delta_1 + \t \Delta_2
\ed
where $\Delta_1$ and $\Delta_2$ are random variables which are
independent of each other and of $y$ and $\t$. Remarkably, they also
do not depend on any properties of the noisy perceptron teacher:
$\Delta_1$ is simply Gaussian with zero mean and variance $Q-R^2$,
while the distribution of $\Delta_2$ follows from the characteristic
function $\bra \exp(-i \hat\Delta \Delta_2)\ket =
\exp(\chi(\hat\Delta))$. All non-Gaussian features of the student
field distribution are encoded in $\Delta_2$. Because $\chi(\cdot)$ is
inversely proportional to $\alpha$, the size of the training set, it
is immediately obvious how the student field distribution recovers its
Gaussian form for $\alpha\to\infty$.

Using the fact that $y$ is Gaussian with zero mean and unit variance,
the training error $E_{\rm {tr}}$ and student field probability
density $P_t(x)$ follow from~(\ref{PS1}) as
\be
E_{\rm {tr}}=\int\! dx \,Dy\sum _{\t =\pm 1}
\theta (-x\t )P_t(x|y,\t )P(\t|y)
\label{PS2}
\ee
\be
P_t(x)=\int\! Dy\sum _{\t =\pm 1}
P_t(x|y,\t )P(\t |y)
\label{PS3}
\ee
in which $Dy=(2\pi)^{-\frac{1}{2}}e^{-\frac{1}{2}y^2}dy$.  We note
again that the dependence of $E_{\rm {tr }}$ and $P_t(x)$ on the
specific noise model -- for a given value of $\rho$ -- arises solely
through $P(\t |y)$. We remind the reader that this teacher output
probability is given by~(\ref{flip}),
\bd
P(\t|y)=(1-\lambda)\ \theta(\t y) + \lambda\theta(-\t y)
\ed
for the case of output noise, while for weight noise~(\ref{gauss2}) implies
\bd
P(\t |y)=\frac {1}{2}[1+\t \, \rm {erf}(y/\sqrt {2}\Sigma)].
\ed

\begin{figure}[t]
\begin{center}
\vspace*{-0.5cm}
\hspace*{0.3cm}
\epsfig{file=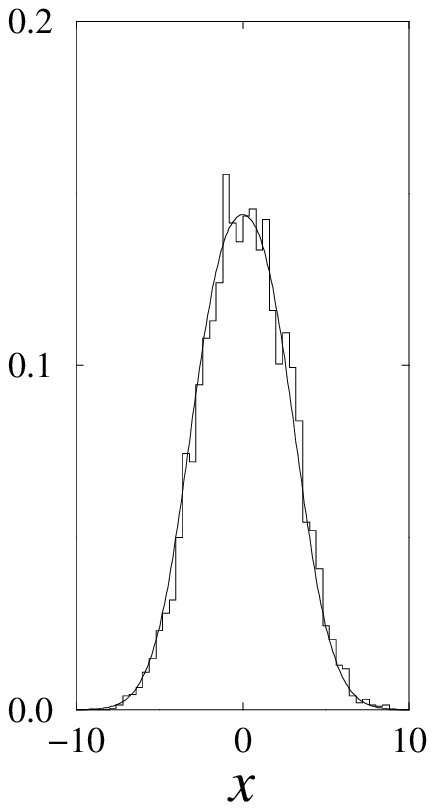,width=4cm}\hspace*{-0.25cm}
\epsfig{file=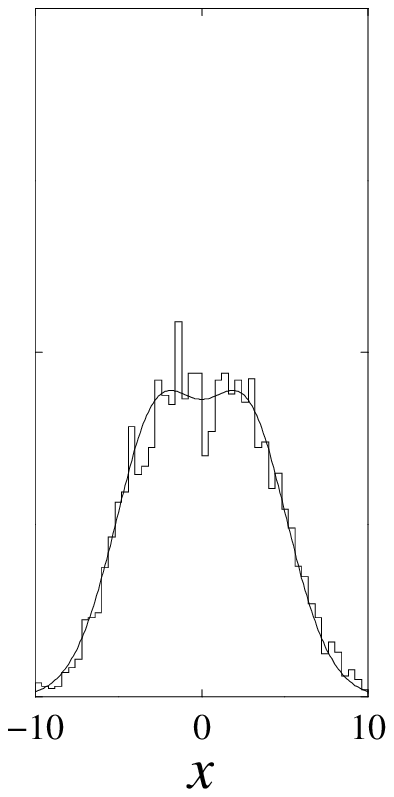,width=4cm}\hspace*{-0.25cm}
\epsfig{file=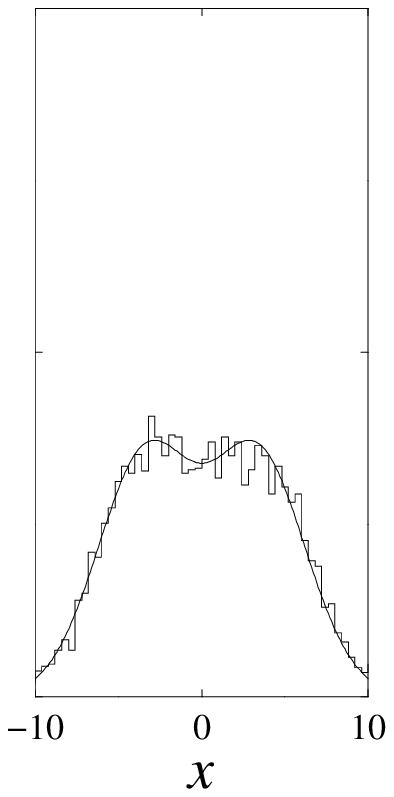,width=4cm}\hspace*{-0.25cm}
\epsfig{file=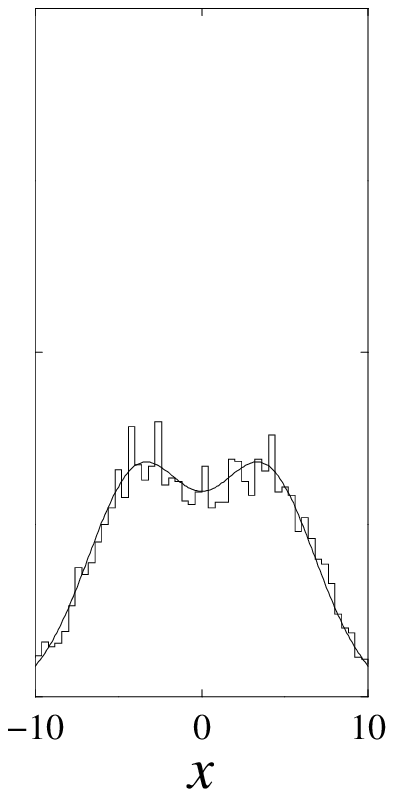,width=4cm}
\end{center}
\vspace*{-0.8cm}
\caption{Student field distribution $P_t(x)$ observed during on-line
Hebbian learning with output noise of strength $\lambda=0.2$, at
different times (from left to right: $t=1,2,3,4$), for training set
size $\alpha=\frac{1}{2}$, learning rate $\eta=1$, and weight decay
$\gamma=\frac{1}{2}$, with initial conditions $Q_0=1$ and $R_0=0$.
Histograms: distributions as measured in numerical simulations of an
$N=10,\!000$ system. Solid lines: predictions of the theory.}
\label{fig:flipfields}
\end{figure}

\section{Comparison with Numerical Simulations}

\begin{figure}[th]
\begin{center}
\vspace*{-0.5cm}
\epsfig{file=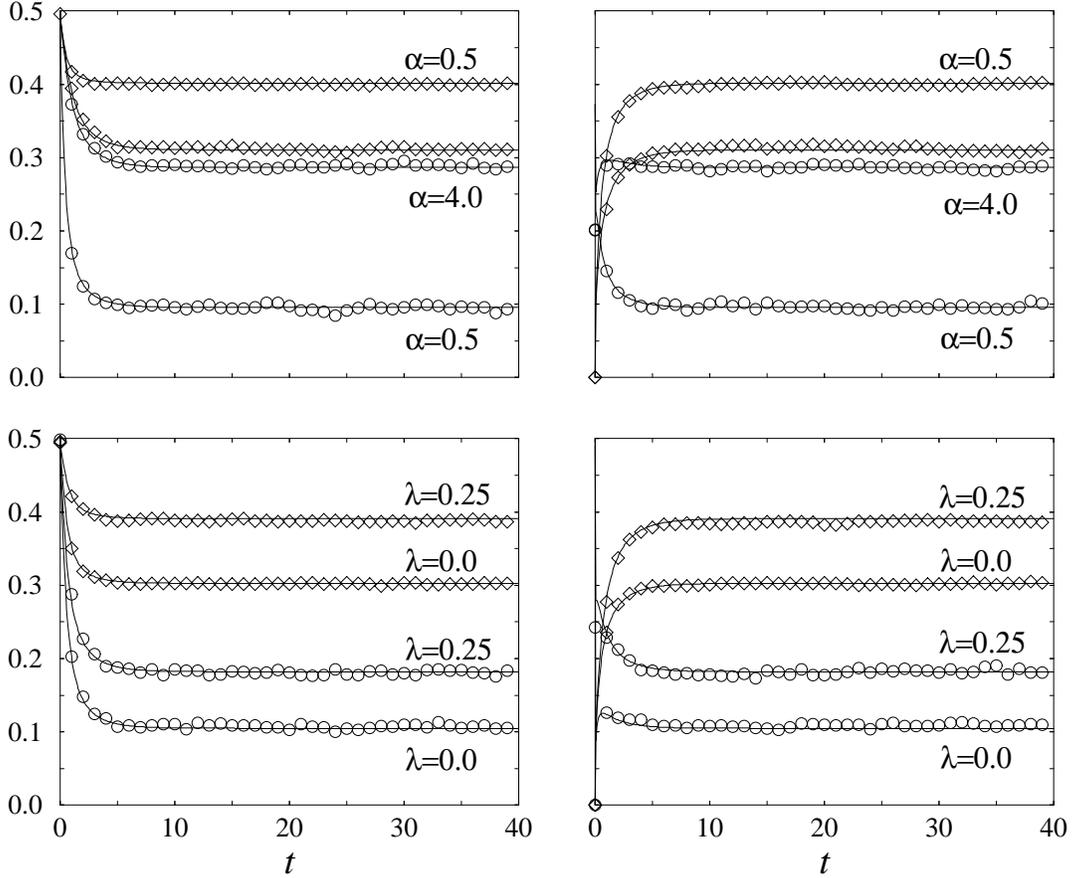}
\end{center}
\vspace*{-0.5cm}
\caption{Generalization errors (diamonds/lines) and training errors
(circles/lines) as observed during on-line Hebbian learning from a
teacher corrupted by output noise, as functions of time.  Upper two
graphs: noise level $\lambda=0.2$ and training set size
$\alpha\in\{0.5,4.0\}$ (initial conditions: upper left, $\egn=0.5$;
upper right: $\egn=0$).  Lower two graphs: $\alpha=1$ and
$\lambda\in\{0.0,0.25\}$ (lower left, $\egn=0.5$; lower right,
$\egn=0$).  Markers: simulation results for an $N=5,\!000$ system.
Solid lines: predictions of the theory. In all cases $Q_0=1$, learning
rate $\eta=1$ and weight decay $\gamma=0.5$.}
\label{fig:fliperrors}
\end{figure}

\begin{figure}[t]
\begin{center}
\vspace*{-0.5cm}
\hspace*{0.3cm}
\epsfig{file=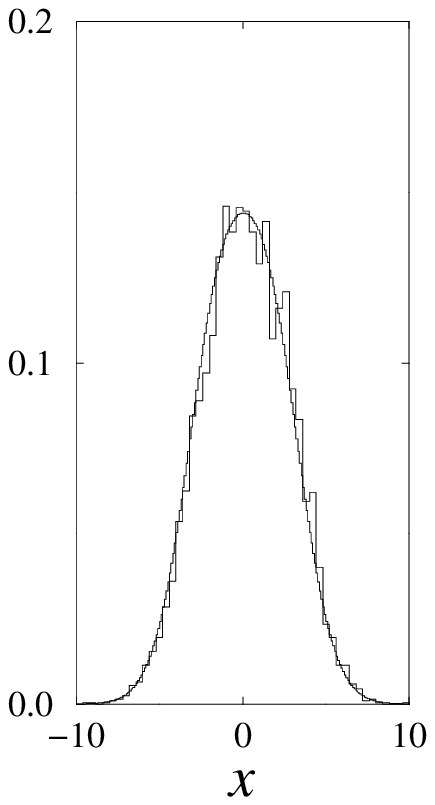,width=4cm}\hspace*{-0.25cm}
\epsfig{file=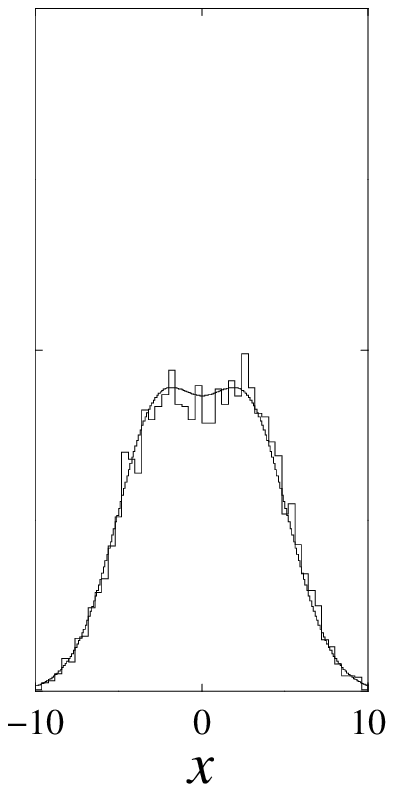,width=4cm}\hspace*{-0.25cm}
\epsfig{file=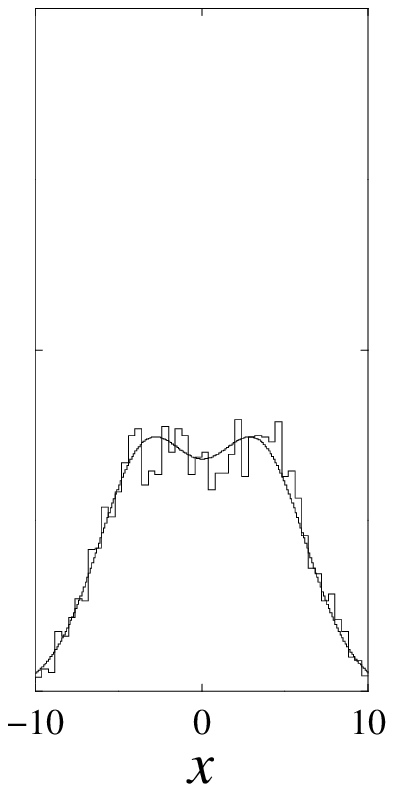,width=4cm}\hspace*{-0.25cm}
\epsfig{file=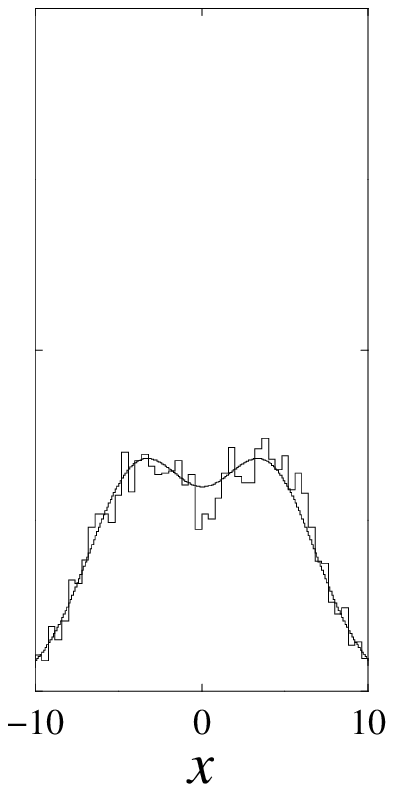,width=4cm}
\end{center}
\vspace*{-0.8cm}
\caption{Student field distribution $P_t(x)$ observed during on-line
Hebbian learning with Gaussian weight noise of effective error
probability $\lambda_{\rm eff}=0.2$ (compare
eq.~(\protect\ref{lambda_eff})), at different times (from left to
right: $t=1,2,3,4$), for training set size $\alpha=\frac{1}{2}$,
learning rate $\eta=1$, and weight decay $\gamma=\frac{1}{2}$, with
initial conditions $Q_0=1$ and $R_0=0$.  Histograms: distributions as
measured in numerical simulations of an $N=10,\!000$ system. Solid
lines: predictions of the theory. See appendix for further discussion
of the close similarities with figure~\protect\ref{fig:flipfields}.
}
\label{fig:gausfields}
\end{figure}

\begin{figure}[t]
\begin{center}
\vspace*{-0.5cm}
\epsfig{file=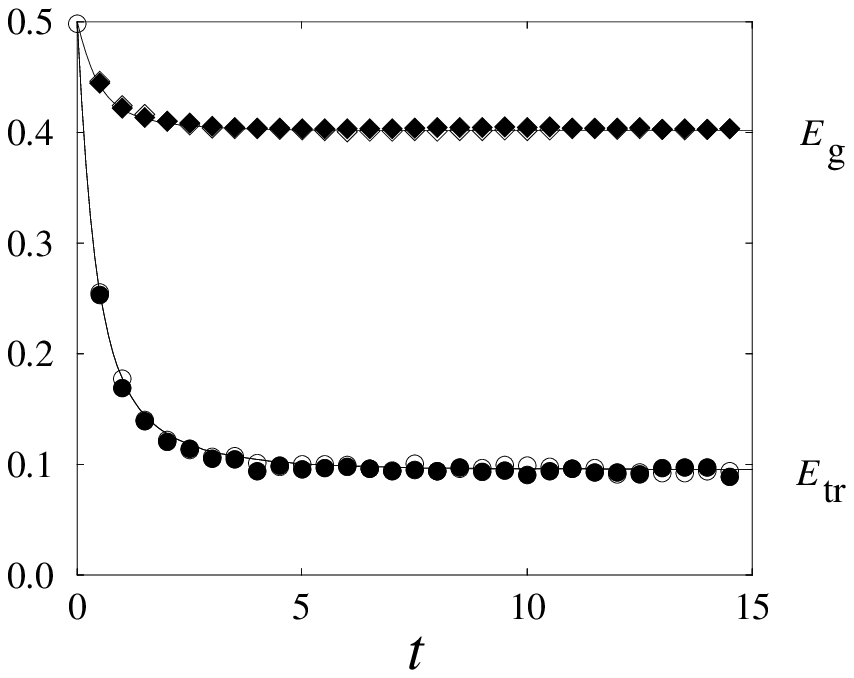}
\end{center}
\vspace*{-0.5cm}
\caption{Comparison between output noise and Gaussian weight noise,
with parameters such that both cases have identical effective error
probability $\lambda_{\rm eff}=0.2$.  Open diamonds (output noise) and
filled diamonds (weight noise): generalization errors as observed in
numerical simulations, as functions of time.  Open circles (output
noise) and filled circles (weight noise): training errors as observed
in numerical simulations, as functions of time.  In all cases training
set size $\alpha=0.5$, learning rate $\eta=1$, weight decay
$\gamma=0.5$, initial conditions $Q_0=1$ and $\egn=0.5$, and system
size $N=10,\!000$.  Solid lines: theory (which here predicts identical
generalization errors and virtually identical training errors).}
\label{fig:gaussdata}
\end{figure}

From the theoretical point of view,
equations~(\ref{PS1},\ref{PS2},\ref{PS3}) constitute the clearest
expression of our results on the joint field distribution since the
dependence of the distribution on the given noise has been separated
out in a transparent manner.  However, we have found that another
equivalent formulation can be useful from the point of view of
numerical computations.  This is detailed in the appendix.

It will be clear that there is a large number of parameters that one
could vary in order to generate different simulation experiments with
which to test our theory. Here we have to restrict ourselves to
presenting a number of representative results.  Figure
\ref{fig:flipfields} shows, for the output noise model, how the
probability density $P_t(x)$ of the student fields $x=\bJ \cdot\bxi $
develops in time, starting as a Gaussian distribution at $t=0$
(following random initialization of the student weight vector) and
evolving into a highly non-Gaussian bi-modal one.  Figure
\ref{fig:fliperrors} compares our predictions for the generalization
and training errors $\eg$ and $E_{\rm {tr}}$ with the results of
numerical simulations (again for teachers corrupted by output noise)
for different initial conditions, $\egn=0$ and $\egn=0.5$, and for
different choices of the two most important parameters $\lambda$
(which controls the amount of teacher noise) and $\alpha$ (which
measures the relative size of the training set).  The system is found
to have no persistent memory of its past (which will be different for
some other learning rules), the asymptotic values of $\eg$ and $E_{\rm
{tr}}$ being independent of the initial student vector%
\footnote{%
In the examples shown, $\eg$ is always larger than $E_{\rm
{tr}}$. However, this is not true generally: We are measuring the
generalization error $\eg$ with respect to the {\em clean} teacher,
whereas the (training) examples that determine the training error
$E_{\rm t}$ are {\em noisy}. Thus, under certain circumstances, $E_{\rm
t}$ can be larger than $\eg$. A trivial example is the case of an
infinite training set ($\alpha\to\infty$) without weight decay
($\gamma=0$). From~(\protect\ref{eginf}), $\eg$ then tends to zero for
long times $t$, while the training error will approach $E_{\rm
t}=\lambda_{\rm real}$, which is nonzero for a noisy teacher.
A generalization error relative to the noisy teacher can also be
defined in our problem; it turns out to be $\eg\mbox{(noisy)} = \{1-\langle
\ta(y)\,\mbox{erf}(yR[2(Q-R^2)]^{-1/2}) \rangle \}/2$.
}%

Figure \ref{fig:gausfields} shows the probability density $P_t(x)$ of
the student fields $x=\bJ \cdot \bxi $ for the Gaussian weight noise
model, with effective error probability $\lambda_{\rm eff}$ chosen
identical to the error probability used to produce the corresponding
graphs \ref{fig:flipfields} for output noise.  Finally we show in
figure \ref{fig:gaussdata} an example of a comparison between the
error measures corresponding to teachers corrupted by output noise and
teachers corrupted by Gaussian weight noise, both with identical
effective output noise probability $\lambda_{\rm eff}=0.2$. Here our
theory predicts both noise types to exhibit identical generalization
errors and almost identical training errors (with a difference of the
order of $10^{-4}$, see the appendix) at any time.  These predictions
are borne out by the corresponding numerical simulations (carried out
with networks of size $N=10,\!000$).  We conclude from these figures
that in all cases investigated the theoretical results give an
extremely satisfactory account of the numerical simulations, with
finite size effects being unimportant for the system sizes considered.

Careful inspection shows that for Hebbian learning there are no true
overfitting effects, not even in the case of large $\lambda$ and small
$\gamma$ (for large amounts of teacher noise, without regularization
via weight decay). Minor finite time minima of the generalization
error are only found for very short times ($t<1$), in combination with
special choices for parameters and initial conditions.  For
time-dependent learning rates, however, preliminary work indicates
that overfitting can occur quite generically.

\section{Discussion}

Starting from a microscopic description of Hebbian on-line learning in
perceptrons with restricted training sets, of size $p=\alpha N$ where
$N$ is the number of inputs, we have developed an exact theory in
terms of macroscopic observables which has enabled us to predict the
generalization error and the training error, as well as the
probability density of the student local fields, in the limit
$N\to\infty$.  Our results are found to be in excellent agreement with
numerical simulations, as carried out for systems of size $N=5,\!000$
and $N=10,\!000$, and for various choices of the model parameters,
both for teachers corrupted by output noise and for teachers corrupted
by Gaussian input noise.  Generalizations of our calculations to
scenarios involving, for instance, time-dependent learning rates or
time-dependent decay rates are straightforward. Closer analysis of the
results for these cases, and for more complicated teachers such as
noisy `reversed wedges', may be an issue for future work.

Although it will be clear that our present calculations cannot be
extended to non-Hebbian rules, since they ultimately rely on our
ability to write down the microscopic weight vector $\bJ$ at any time
in explicit form~(\ref{b2}), they do indeed provide a significant
yardstick against which more sophisticated and more general theories
can be tested. In particular, they have already played a valuable role
in assessing the conditions under which a recent general theory of
learning with restricted training sets, based on a dynamical version
of the replica formalism, is exact~\cite{coolensaad,coolensaad2}.

{\bf Acknowledgments:} PS is grateful to the Royal Society for
financial support through a Dorothy Hodgkin Research Fellowship.

\clearpage

\appendix
\section{Evaluation of the Field Distribution and Training Error}

In this appendix, we give alternative forms of our main
results~(\ref{PS1},\ref{PS2},\ref{PS3}) for the joint field
distribution and training error that are more suitable for numerical
work. For this purpose, it is useful to shift attention from the noisy
teacher output $\t$ to the corrupted teacher field $z$ that produces
it; the two are linked by $\t=\sgn(z)$. This is entirely natural in
the case of Gaussian weight noise. As discussed after
eq.~(\ref{gauss}), $z$ then differs from the clean teacher field $y$
by an independent zero mean Gaussian variable with variance
$\Sigma^2$; explicitly, one has the conditional distribution
\bd
P(z|y)= \frac{1}{\sqrt{2\pi\Sigma^2}}\, e^{-(z-y)^2/2\Sigma^2} \qquad
\mbox{(Gaussian weight noise).}
\ed
The case of output noise can be treated similarly, by assuming that
$z$ is identical to $y$ with probability $1-\lambda$, but has the
opposite sign with probability $\lambda$:
\be
P(z|y)=(1-\lambda)\, \delta(z-y) + \lambda\,\delta(z+y) \qquad
\mbox{(output noise).}
\label{zflip}
\ee
We now consider the joint distribution $P_t(x,y,z)$. It can be
derived by complete analogy with the calculation in
section~\ref{sec:distribution}. For the conditional distribution of
$x$, one finds that
\bd
P_t(x|y,z)=P_t(x|y,\sgn(z)).
\ed
Intuitively, this follows from the fact that during learning, the
student only ever sees the noisy teacher output $\sgn(z)$, but not the
corrupted field $z$ itself; the student field $x$ can therefore depend
on $z$ only through $\sgn(z)$.  Multiplying by the joint distribution
of $y$ and $z$, and using the result~(\ref{PS1}), one thus finds, for
the case of output noise,
\bd
P_t(x,y,z) = \left[(1-\lambda)\, \delta(z-y) +
\lambda\,\delta(z+y)\right] 
\frac {e^{-\frac {1}{2}y^2}}{\sqrt {2\pi }}
\int\! \frac {d\hx }{2\pi }\, e^{-\frac {1}{2}{\hx }^2(Q-R^2)
+i\hx (x-yR)+\chi (\hx\sgn(z))}
\ed
with the marginal distribution
\be
P_t(x,z) = \frac {e^{-\frac {1}{2}z^2}}{\sqrt {2\pi }}
\int\! \frac {d\hx }{2\pi }\, e^{-\frac {1}{2}{\hx }^2(Q-R^2)
+i\hx x +\chi (\hx\sgn(z))} 
\left[(1-\lambda)e^{-i\hx zR}+
\lambda e^{i\hx zR}\right].
\label{output_marginal}
\ee
The corresponding expressions in the case of Gaussian weight
noise read
\bd
P_t(x,y,z)=\frac {1}{2\pi \Sigma }
\int\! \frac {d\hx }{2\pi }\,
e^{-\frac {1}{2}{\hx ^2}(Q-R^2)+i\hx (x-Ry)
+\chi (\hx \sgn(z))
-[z^2-2yz +y^2(1+\Sigma ^2)]/(2\Sigma ^2)}
\ed
and
\be
P_t(x,z)=
\frac {e^{-\frac {1}{2}z^2/(1+\Sigma ^2)}}{\sqrt {2\pi (1\plus\Sigma
^2)}} \int\! \frac {d\hx }{2\pi }\, e^{-\frac {1}{2}\hx
^2[Q-R^2/(1+\Sigma ^2)]+i\hx [x-Rz/(1+\Sigma ^2)]+\chi (\hx \sgn (z))}
\ .
\label{bb55}
\ee
In both cases, the training error and the probability distribution of
the student field $x$ are then determined by
\bd
E_{\rm {tr}}=\int\! dx\,dz\,\theta (-xz)\,P_t(x,z)
\qquad
P_t(x)=\int\! dz\,P_t(x,z)
\ed
respectively. For a numerical computation of these two quantities, it
is imperative to further reduce the number of integrations
analytically, which turns out to be possible. In the following, we
drop the time subscript $t$ on all distributions to save notation.

First we deal with the case of output noise. In the marginal
distribution~(\ref{output_marginal}), we make the change of variable
$\hat{x}=k\sgn(z)$ to get
\bd
P(x,z)=\frac {e^{-\frac {1}{2}z^2}}{\sqrt{2\pi}}\,
\int\frac{dk}{2\pi}\,e^{ -\half k^2(Q-R^2)+{\chi }(k)+ik x\sgn (z)}
\left\{ (1\minus \lambda )e^{-ik|z|R }+\lambda e^{ik|z|R}\right\}.
\ed
The training error is
\bd
E_{\rm tr}=\int\! dx\,dz\, P(x,z)\,\theta (-xz) 
=\int _0^\infty\!dx\, \left[P_{+}(-x)+P_{-}(x)\right]
\ed
where
\be
P_{\pm }(x)=\int dz\, P(x,z)\,\theta (\pm z)
=\half \int \frac {dk}{2\pi}\, Dz\, e^{ -\half k^2(Q-R^2) + {\chi }(k)\pm
ik x}\left\{
(1\minus \lambda)e^{-ik|z|R} +
\lambda e^{ik|z|R }
\right\}
\label{ss5}
\ee
We see that $P_{+}(x)=P_{-}(-x)\equiv \Pi (x)$. In terms of $\Pi (x)$
we have the formulae
\be
P(x)=\Pi (x)+\Pi (-x)
\qquad
E_{\rm tr}=2\int _0^\infty\! dx \, \Pi (-x) 
\label{ss6}
\ee
The function $\Pi(x)$ can be further simplified by decomposing $\chi$
into its real ($\chi_{\rm r}=\RE(\chi)$) and imaginary ($\chi_{\rm i}=\IM(\chi)$)
parts:
\begin{eqnarray}
\lefteqn{
\Pi (x) =  \int \frac {dk}{4\pi}\, Dz\, e^{ -\half k^2(Q-R^2) +
\chi(k) + ikx}\left\{ 
(1-\lambda)e^{-ik|z|R}+
\lambda e^{ik|z|R}
\right\}
}
\nonumber\\
& \!=\! & \!\!\! \int \frac {dk}{4\pi}\, Dz\, e^{ -\half k^2(Q-R^2) +
{\chi_{\rm r} }(k)} \left\{ 
(1\minus\lambda )\cos [{\chi_{\rm i} }(k) \plus k(x\minus R|z|)] +
%
%
%
\lambda \cos [ {\chi_{\rm i} }(k) \plus k(x\plus R|z|)]
\right\}
\nonumber\\
& \!=\! & \!\!\! \int \frac {dk}{4\pi}\, e^{ -\half Q k^2 + {\chi_{\rm r}
}(k) } \left\{ \cos [ {\chi_{\rm i} }(k)\plus k x]
%
%
%
+(1\minus 2\lambda)\sin [ {\chi_{\rm i} }(k)\plus k x]\,G(k R) \right\}
\label{Pi_of_x}
\end{eqnarray}
in which 
\be
G(\Lambda )=e^{\frac {1}{2}{\Lambda }^2}\int\! Dz\, \sin (\Lambda |z|)
=\frac {\Lambda }{\sqrt {\pi }}\ {_{1}}F_{1}\left(\frac {1}{2};\frac
{3}{2};\frac {1}{2}{\Lambda }^2\right)
\label{ss20}
\ee
and ${_{1}}F_{1}(\ldots)$ is the degenerate hypergeometric function
(see~\cite{GS}, page 1058).  From equation~(\ref{ss6}) we now
immediately obtain our final result for the student field
distribution:
\be
P(x) = \int \frac {dk}{2\pi}\, 
e^{-\frac {1}{2}Q k^2 + {\chi_{\rm r} }(k )} \cos (k x)
%
%
%
\left\{\cos [ {\chi_{\rm i} }(k )]
+(1\minus 2\lambda)G(k R)\sin [ {\chi_{\rm i} }(k )]\right \}
\label{ss10}
\ee
To further simplify the expression~(\ref{ss6}) for the training error,
we write
\bd
E_{\rm tr}=
\lim _{L\rightarrow \infty}2\int_{-L}^0\! dx\,\Pi (x)= 
2\lim _{L\rightarrow \infty }I(L)\label {ss11}
\ed
where, from~(\ref{Pi_of_x})
\begin{eqnarray*}
I(L) & = & \int \frac {dk}{4\pi}\, e^{-\half Q k^2 + {\chi_{\rm r}
}(k )}\left\{\int _{-L}^0\! dx \, \cos [
{\chi_{\rm i} }(k )\plus k x] 
+ (1\minus 2\lambda) G(k R) \int _{-L}^0\! dx\, \sin [ {\chi_{\rm i}
}(k )\plus k x]\right\}
\end{eqnarray*}
Thus
\bea
I(\infty) & = & - \int \frac {dk}{4\pi k}\, e^{-\half Q k^2 +
{\chi_{\rm r} }(k )}
\left\{(1\minus 2\lambda )G(k R)\cos [ {\chi_{\rm i} }(k 
)]-\sin [ {\chi_{\rm i} }(k )]\right\}+
\nonumber\\
& & \lim _{L\rightarrow \infty }\int \frac {dk}{4\pi k}\, e^{-\half Q k^2 +
{\chi_{\rm r} }(k )}
\left\{\sin [k L \minus  {\chi_{\rm i} }(k )]
+ (1\minus 2\lambda )
G(k R)\cos [k L\minus  {\chi_{\rm i} }(k )]\right\}
\label{ss12}
\eea
The $L$-dependent integral in (\ref {ss12}) can be expressed as a sum
of two integrals, which we consider separately.  In the first part, we
replace $k$ by $k/L$ and obtain
\begin{eqnarray*}
\lefteqn{
\lim _{L\rightarrow \infty }
\int \frac {dk}{4\pi k}\, e^{-\half Q k^2 + {\chi_{\rm r} }(k )}
\sin [k L\minus  {\chi_{\rm i} }(k )]
}
\\
& & = \lim _{L\rightarrow \infty }\int\! \frac {dk}{4\pi k}
\,e^{-\half Q (k/L)^2 + {\chi_{\rm r} }(k /L)}
\sin [k \minus  {\chi_{\rm i} }(k /L)]
=\int\! \frac {dk}{4\pi k}\,\sin (k )=\frac {1}{4}\ .
\end{eqnarray*}
Secondly we need to consider the behaviour of
\be
\int\! \frac {dk}{4\pi k}\, e^{-\half Q k^2 + \chi_{\rm r}(k)}
\cos [k L\minus  {\chi_{\rm i} }(k )] G(k R)
\label{ss14}
\ee
in the limit $L\rightarrow \infty$.  We set $u=kR$ and note that,
because $Q\geq R^2$, one has $e^{-\half Q k^2} \leq e^{-\half u^2}$;
furthermore,
\bd
\left| e^{-\frac {1}{2}u^2}G(u)u^{-1} \right|
=\left|\int Dz\, |z|\frac {\sin (|uz|)}{|uz|} \right| \leq \int Dz\,
|z|=\sqrt {\frac {2}{\pi }}\ .
\ed
Finally, $ {\chi }(k )$ is independent of $L$ and is bounded as a
function of $k$; in fact, from~(\ref{chi_def}), $| {\chi }(k )|\leq
2\alpha ^{-1}t$. It follows by an application of the Riemann-Lebesgue
Lemma (see e.g.~\cite{TITCH}) that the integral~(\ref{ss14}) tends to
zero as $L\rightarrow \infty$.  We conclude that for output noise the
training error is given by
\be
E_{\rm tr}=
\frac {1}{2}-\int\! \frac {dk}{2\pi k}\, e^{-\half Q k^2 + \chi_{\rm r}(k)}
\left\{
(1\minus 2\lambda ) G(k R) \cos [ {\chi_{\rm i} }(k )] 
-\sin [  {\chi_{\rm i} }(k )] 
\right\}
\label{ss15}
\ee
where $G(\ldots)$ is defined by~(\ref{ss20}).

\begin{figure}[t]
\begin{center}
\epsfig{file=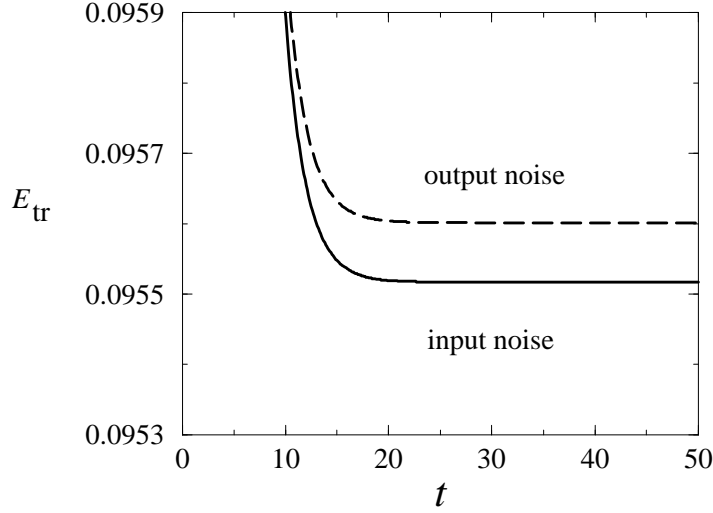}
\end{center}
\vspace*{-0.5cm}
\caption{Characteristic example of theoretical predictions for the
training error $E_{\rm tr}$ for two noisy teachers with identical
effective error probability $\lambda_{\rm eff}=0.2$. Dashed line:
output noise; solid line: Gaussian weight noise. Parameters:
$\alpha=\gamma=0.5$, $Q_0=\eta=1$, $\egn=0.5$.}
\label{fig:difference}
\end{figure}

The procedure for Gaussian weight noise is similar to that of output
noise.  We start from equation~(\ref{bb55}) and define
\bd
\rrt=R/\sqrt {1\plus \Sigma^2}.
\ed
Upon defining $\hat{x}=k\sgn(z)$ in~(\ref{bb55}), replacing $z$ by
$z/\sqrt{1+\Sigma^2}$, and continuing in the same notation as for
output noise, we find
\be
P_{\pm}(x)=\half \int \frac {dk}{2\pi}\, Dz \,
e^{-\frac {1}{2}k^2(Q-\rrt^2) + 
 {\chi }(k)\pm ik x -ik \rrt |z|}
\label{tt5}
\ee
Since~(\ref{tt5}) can be obtained from~(\ref{ss5}) by putting $\lambda
\!\to\! 0$ and $R\!\to\! \rrt$, we immediately
obtain for the student field distribution and the training error,
respectively [see equations~(\ref{ss10}) and~(\ref{ss15})],
\be
P(x)=\int \frac {dk}{2\pi}\, e^{-\half Q k^2 +
{\chi_{\rm r} }(k)} \cos (k x)\left\{\cos
[ {\chi_{\rm i} }(k)]\plus G(k\rrt)\sin [ {\chi_{\rm i} }(k)]\right\}
\label{tt10}
\ee
\be
E_{\rm tr}=
\frac {1}{2}-\int\! \frac {dk}{2\pi k}\, e^{-\half Q k^2 + 
\chi_{\rm r}(k)}
\left\{
G(k \rrt) \cos [ {\chi_{\rm i} }(k )]
- \sin [  {\chi_{\rm i} }(k )] 
\right\}
\label{tt15}
\ee
In particular, we can now calculate the student field distribution and
the training error for both output noise and Gaussian weight noise,
with noise levels such that in both cases $\lambda_{\rm eff}=\lambda$.
This guarantees that, at any time, $Q$, $R$ and $\eg$ will have the
same values in both cases; it also implies
$\rrt=R/\sqrt{1+\Sigma^2}=R(1-2\lambda)$. We then obtain
from~(\ref{ss10},\ref{ss15},\ref{tt10},\ref{tt15}) very similar
expressions:
\begin{eqnarray*}
P^{\rm out}(x) & = & \int\!\frac{dk}{2\pi}\, e^{- \frac {1}{2}Qk^2 +
\chi_{\rm r}(k)} \cos(kx)
\left\{
\cos[ {\chi_{\rm i} }(k)] + (1 - 2\lambda) G(kR) 
\sin [ {\chi_{\rm i} }(k)] \right\}
\label{eq:flipfields}
\\
P^{\rm gau}(x) & = & \int\!\frac{dk}{2\pi}\, e^{ -\frac {1}{2}Qk^2 + 
{\chi_{\rm r}}(k)}\cos(kx)
\left\{
\cos[ {\chi_{\rm i} }(k)] + 
G[(1\minus 2\lambda)kR] \sin [ {\chi_{\rm i} }(k)] \right\}
\label{eq:gausfields}
\end{eqnarray*}
and
\begin{eqnarray*}
E^{\rm out}_{\rm tr} & = &
\frac {1}{2}-\int\! \frac {dk}{2\pi k}\, e^{ -\frac {1}{2}Qk^2 +
{\chi_{\rm r}}(k)}
\left\{
(1 - 2\lambda) G(kR) \cos [ {\chi_{\rm i} }(k)]
- \sin[ {\chi_{\rm i} }(k)]
\right\}
\label{eq:fliperror}
\\
E^{\rm gau}_{\rm tr} & = &
\frac {1}{2}-\int\! \frac {dk}{2\pi k}\,e^{ -\frac {1}{2}Qk^2 + 
{\chi_{\rm r}}(k)}
\left\{
G[(1\minus 2\lambda)kR] \cos[ {\chi_{\rm i} }(k)]
- \sin[ {\chi_{\rm i} }(k)]
\right\}
\label{eq:gauserror}
\end{eqnarray*}
Provided parameters are chosen such that the effective error
probabilities are identical, the differences between output noise and
Gaussian weight noise are restricted to the positioning of the factor
$1\minus 2\lambda$ relative to the integral $G(\ldots)$, with
manifestly identical expressions for $\lambda=0$ and
$\lambda=\frac{1}{2}$ (as it should be).  As a result the resulting
curves for field distributions and training errors are found to be
almost identical; figure~\ref{fig:difference} shows a typical example.

\end{document}